\documentclass[12pt,preprint]{aastex}%
\usepackage{natbib}
\usepackage{amsmath}
\usepackage{graphicx}%
\usepackage{amsfonts}%
\usepackage{amssymb}

\shorttitle{Precession}\shortauthors{Meirelles }

\begin{document}

\title{An Alternative Look To Precession In Accretion Disks}

\author{Cesar Meirelles Filho}

\affil{Instituto de Astronomia, Geof\'{\i}sica e de Ci\^{e}ncias Atmosf\'{e}ricas \\
Universidade de S\~ao Paulo \\R. do Mat\~ao, 1226, 05508-090 S\~ao Paulo, SP, Brasil}

\email{cmeirelf@astro.iag.usp.br}

\begin{abstract}

We have considered precession in accretion disks in which a second moment of inertia relative to an axis perpendicular to the axis of rotation may be very 
important. This formalism, that takes into account the precession contribution to the angular momentum, is based on the existence of a parameter $\it p$ which  determines three characteristic densities resulting from the averaging process and imposes 
constraints on the actual disk density. It is shown that the precession velocity will lie in a three branch solution, and depends on how large is the disk actual
 density as compared to the characteristic densities. Besides the large spread on the solution for the precession velocity, depending on the density strength,
 it may be prograde and retrograde. It is shown that the keplerian thin disk, with very large density values compared to characteristic ones  , only precesses very far away from the primary object,
 which implies very large precession periods. For other models, the disk will thicken, with large deviations from the keplerian approximation. Constraints on 
the density only will be effective for very large values of the ratio ${{\dot M} \over M_{p}}$, respectively, accretion rate and mass of the primary. Under this condition, the structure of the precessing region is 
found. Lower bounds on the precession period are found for not so large values of this ratio. Deviations from the mean precessional motion are considered. 
It is shown that these deviations result in periodic motions as long as the time scales associated to them are comparable to the remaining time scales. 
Otherwise, they result in misalignment motions, forcing the plane of the disk to become normal to the orbital plane of the secondary.

\end{abstract}

\section{Introduction}
       
	Precessional activity has long been invoked to explain time variations occurring in the spectra of  galactic X-ray binaries 
like LMC X-4, Her X-1, SS 443, Cygnus X-1, etc. Observational evidences give support to the idea that these systems, and quite a lot 
of suspected others,  consist of a disk-like system plus a third body. This is very suggestive  of precession as  
being the effect  due to the perturbing torque of a distant companion star in the the disk or, even,  as the result in the disk of the 
secondary star precessing in the tidal field of the central compact object. As a matter of fact, the scenario is not unique, and according to \citet{prie87}, 
there are, at least , five possibilities. This is so because there is uncertainty concerning the reason why the disk precesses, where and the extent 
on it that precesses. The possibilities  explicitly dependent upon tidal torques under which the disk will precess  can be summarized by the following
 two scenarios:   

a- the disk has a permanently tilted edge that freely precesses in the tidal field of the companion star. This kind of scenario has been first exploited
 by \citet{ka73}, trying to explain the 35 day period on the light curve of Her X-1.  This same model was used by \citet{ka80}  to explain observational 
features in SS 443;

b- the companion star precesses in the tidal field of the central compact star, slavishly followed by the the outer tilted disk. This kind of scheme was
 proposed by \citet{rob74}, also to explain the 35 day period observed in Her X-1. 

	However, precession is not a exclusiveness of galactic X -ray binaries sources, its presence being suspected in a lot of other astrophysical systems.
 Jet structures observed in AGN, and also in protostars seem to be associated with a precessing disk ( \citet{li99}). In the context of disks around supermassive objects, the reason why the disk precesses is still more unclear. 
 \citet{ka97}  suggested that observational data in OJ 287 could be explained by a tidal torque due to the presence of another massive companion star, in a way 
similar to that employed to Her X-1 and SS 443.  \citet{prin96} and \citet{prin97} suggests precession as a result of  a warping instability caused by the central
 source irradiation.  \citet{rom00} argue, in analogy with \citet{ka82}and \citet{ka97},, aiming to explain the pattern of ejection observed in the Quasar 3C 273, 
 that  precession is due to a massive secondary black hole tidal torque on the disk,a model, subsequently,  applied to 3C 345 by \citet{ca02}.

	It should be argued, however, that precession is a kind of solid-like response of the disk. A fluid-like response would be a more appropriate one
  for a fluid system like the disk. Besides,   in the case of fluid-like response, the tidal torque acts as a perturbation, which main consequence is the
 generation of different sort of waves. Quite a variety of physical issues, suitable for a wave approach, such as several tidally driven instabilities, disk
 tidal deformation (warping), horizontally  and vertically driven resonances, angular momentum transport, and so on (\citep{lu91, lua92, lu93, olga02, olgb03,
 vis92}) have been tackled quite successfully.  In addition, it has been shown  that, under this approach,  for disks in systems with extreme mass 
 ratio, precession may occur due to a coupling between an eccentric instability and Lindblad resonances (\citet{lua92}).  \citet{pat95},
 extending \citet{lua92}    result, argue that the axisymmetric part of the tidal potential is related to the disk solid-like response, while
 the disk fluid-like response is related to the non-symmetric part of the potential. In these works, the the disk solid body-like response comes as an integrabil;ity 
condition.

	\citet{ka82} proposal to compute frequencies treats the disk as a ring, with no allowance given for the width in the plane of the ring, nor to the thickness 
perpendicular to that plane. By employing a suitable averaging procedure, he obtains for the for the precession angular velocity

$${\Omega}_{p}=-\frac{3}{4} \, \frac{{{\omega}_{s}}^2}{{\Omega}_{d}} \, cos{\delta},$$

where ${\omega}_{s}$ is the keplerian angular velocity of the secondary, ${\Omega}_{d}$ is the ring angular velocity, and $\delta$ is the inclination angle.
	This formalism, besides treating the disk as a rigid body, not taking into account its structure, assumes the disk angular velocity parallel to the 
disk  angular momentum . Though keeping $\vec{{\Omega}_{p}}$ parallel to ${\vec L}$, assuming thin disks obeying polytropic equation of state, the structure 
has been considered by \citet{pat95}, \citet{laral96}, \citet{larp97} and \citet{lar97}. Decomposing the secondary tidal potential into odd and even z parts, with subsequent 
Fourier decomposition, and arguing that the odd axisymmetric part is responsible for rigid precession, these authors,assuming ${\ell} \over r$ constant, obtain

$${\Omega}_{p}=-\frac{3}{4} \,   \, {\left( \frac{7-2 \, n}{5-n} \right)} \, \frac{{{\omega}_{s}}^2}{{\Omega}_{d}} \, cos{\delta},$$

  as an integrability condition. In this expression, n is the polytropic index and ${\Omega}_{d}$ is the disk angular velocity evaluated at the outer edge of the 
disk, $R_{0}$. For $n=2$, they recover \citet{ka82} result. According to them, the validity of this result holds whenever the disk doesn't thicken, the perturbative tidal 
potential is weak, and, above all, the sound crossing time  is shorter than the precession period.
	It should be argued, however, that the disk angular momentum and the precession velocity are not parallel. Besides, heat generation, cooling and angular momentum 
 transport do occur in accretion disks, a fact that makes not realistic the structure obtained  from a polytropic equation of state. This imposes severe restrictions 
on the model of disk we want to consider.It is reasonable to expect the polytropic index varying from $n=\frac{3}{2}$, for a (monoatomic) gas pressure dominated disk, to $n=3$ 
for a radiation pressure dominated one. Disks with negligible temperature gradients would require  a polytropic index much greater than 3: $\gamma =1$ is appropriate for an isothermal 
atmosphere ( $n \rightarrow \infty $). In that situation, the mass content in the disk, the moment of inertia and the precession velocity are determined by values 
 close to the inner radius rather than close to the outer radius, the disk structure being very susceptible to relativistic corrections and gravitational radiation 
losses    may become important.  Besides,  if one is interested in a region in parameter space $M-{\dot M}$ where the disk thickens, the dynamics 
can not be dissociated from the structure, as is the case for the keplerian thin disk. In other words, the azimuthal velocity depends on the disk height scale      .
 In addition, for a given tidal potential, its relative strength increases as the disk thickens. For actual radiation pressure dominated, the height scale is constant,
 the polytropic index is negative, implying negative specific heat. Besides this, in this case, one can not satisfy null boundary condition at the surface of the 
disk  Finally, the precession period diminishes relative to the sound crossing time as the disk gets thick. A final possible question concerning the applicability of 
of these results to disks is related to the precession velocity dependence on the outer radius of the disk, which yields proportionality with the mass of the disk. 
Despite implying larger inertia moments, the larger the mass the faster the disk will precess.

	\citet{rom00} work consists essentially of the application of \citet{ka82} and \citet{ka97} to a binary system composed by a disk around a primary supermassive black hole, plus a supermassive secondary orbiting the primary in a keplerian way, not coplanar to the disk. It should be argued, however, that the formulation followed by these authors, besides suffering from the drawbacks, we just mentioned, of any precessing disk model, relies heavily on the above cited authors results for the precession angular 
velocity, which was obtained without considering  the precession contribution to the total angular momentum. Besides, this model is supposed to apply to disks eventually  
dominated by radiation pressure.  In addition, account of the precession contribution to the angular momentum introduces some non-linearity in the problem, which, in turn, introduces conditions on the averaging procedure, 
imposing constraints on the density, on the disk unperturbed angular velocity, on the viscosity parameter as well as on the height scale of the disk. In that case,
 there will be a contribution from the precession velocity to a motion in a plane perpendicular to the plane containing the normal to the disk and 
 the normal to the plane of the orbit of the secondary. This motion, depending on the magnitude of its time scale, will be periodic or not .

	In the following we shall address the question of precession in accretion disks in systems similar to that considered by \citet{rom00}. In our 
formulation, however, we will be concerned to treat the secondary tidal torque on the disk as a perturbation, which means that the secondary never
 crosses the disk, being far away from any point in the disk. Granted this, we will take into account the precession contribution to the total angular momentum
 and will discuss the conditions under which we may take an average of the Euler equations, in such a way as to keep algebraic solutions to the precessional motion.
Deviations from the precessional motion will be treated by solving the equation for the time evolution of $\delta$, the angle that measures the misalignment 
between the normal to the plane of the disk and the normal to the plane of the secondary orbit. The time evolution will be obtained assuming the time scale for this 
motion is large or comparable to the remaining time scales. We obtain results for the precession velocity for different disk models, taking into account deviations from geometrical 
thinness due to a relation between the angular velocity of the disk and the height scale. One of the main results of this work is to show that, if precession
 is to occur in the inner region of disks around a supermassive black hole, the disk is not keplerian, being very thick and luminosity deficient.

\section{The System and the Dynamics } 

The system we consider is a binary one composed by a primary compact object (a black hole), of mass $\it{M_{p}}$, a massive secondary 
object (another black hole), of mass $\it{M_{s}}$, and an accretion disk around the primary one. The secondary is orbiting the primary plus disk 
in a plane inclined of an angle $ \delta$ relative to the normal of the plane of the disk. Working in a reference frame tied to the disk, a point on
it is at a distance ${\it{R}}$ from the center (the primary star) and has coordinates ({\it{x, y, z}}) or ($ {\it{r, \phi, z}}$). 
The line of nodes will be chosen coincident with the x-axis, and the position of the perturber (secondary) will be

$$\vec{d} = d \, {\left( cos{\omega} \, t \, \vec{i} + cos{\delta} \, sin{\omega} \, t \, \vec{j} + sin{\delta} \, sin{\omega} \, t \, \vec{k} \,  \right)} \,\, \eqno(1)  $$

where $\omega$ is the angular velocity of the secondary.

The potential at the point (r, $\phi $, z) is 

$$ \Psi = -\frac{ -GM_{p}} {R} - \frac{G M_{s}}{ | \vec{R}-\vec{d}|} + \frac{G  M_{s}} {d^{3}} . \vec{R}.\vec{d}, \,\,\eqno(2) $$

G being the gravitational constant. The first term is the potential due to the primary, the second is the tidal potential due to the secondary, and the 
last takes into account the acceleration of the coordinate system.
	The tidal force per unit mass will be

$$ f_{r}= -G M_{s} A(r, z, t, {\phi}, d, {\omega}, {\delta})^{-3/2}  C(r, d, t, {\omega}, {\phi}, {\delta}) - {\frac{{G  M_{s}}} d^{2}}  B(t, {\omega}, 
{\phi}, {\delta}),\, \, \eqno(3) $$

 $$ f_{\phi}= G \, M_{s} \, A(r, z, t, {\phi}, d, {\omega}, {\delta})^{-3/2} \,  d \, Q(t, {\omega}, {\phi}, {\delta}), \,\, \eqno(4)  $$

and

$$ f_{z}= -G  M_{s}  A(r, z, t, {\phi}, d, {\omega}, {\delta})^{-3/2} N(z, d, t, {\omega},  {\delta}) - {\frac{{G M_{s}}} d^{2}} \, E(t, {\omega},  
 {\delta}) \,\,  \eqno(5) $$

Where A, B, C, E, N, and Q are given by

$$ A(r, z, t, {\phi}, d, {\omega}, {\delta})= {\left( d^2+r^2+z^2-2 \, r\, d\,  B(t, {\omega}, {\phi}, {\delta})-2 \, z\, d \, E(t, {\omega}, {\delta})\right)}, \,\,\eqno(6) $$

$$ B(t, {\omega}, {\phi}, {\delta})= cos{\omega} \, t \, cos{\phi} + cos{\delta} \, sin{\omega} \, t \, sin{\phi}, \,\,  \eqno(7) $$

$$ C(r, d, t, {\omega}, {\phi}, {\delta})= r - d \, B({\omega}, {\phi}, {\delta}), \eqno(8) $$

$$ E(t, {\delta}, {\omega}, t)= sin{\delta} \, sin{\omega} \, t ,  \eqno(10) $$

$$   N(z, d, t, {\omega},  {\delta})= z - d \, E(t, {\delta}, {\omega}, t),  \eqno(11)  $$

$$ Q(t, {\omega}, {\phi}, {\delta})= - sin{\phi} \, cos{\omega} \, t  + cos{\delta} \, sin{\omega} \, t \, cos{\phi}.  \eqno(12) $$

	To obtain the the precession angular velocity we must take into account its contribution  to the angular momentum. In Fig.1, we present a schematic
 representation of the system we are considering.

where z is the coordinate normal to the plane of the disk and ${\Omega}$ , the precession angular velocity,  makes an angle ${\delta}$ with z. 
Therefore, the angular momentum is no longer parallel to the normal to the plane of the disk, and we may write

$$ \vec{L}= \vec{L_{\parallel}} + \vec{L_{\perp}} $$

$$  = I_{1} \, {\vec{{\omega}_{1}}} + I_{2} \, {\vec{{\omega}_{2}}} ,  \eqno(13)$$

where $I_{1} $ and $I_{2} $ are respectively the moments of inertia relative to an axis normal to the plane of the disk and another that lies in the plane of 
the disk (z=0);  $ {\vec{{\omega}_{1}}} $ and  $ {\vec{{\omega}_{2}}} $ are angular velocities parallel to those axes.
	It is straightforward to see

$$ {\omega}_{1} = {\Omega}_{d} + {\Omega}_{p} \, cos{\delta},  \eqno(14) $$

$$ {\omega}_{2} = {\Omega}_{p} \, sin{\delta} , \eqno(15)$$

${\Omega}_{d} $ being the angular velocity in the disk.

	So, applying these considerations to an element of matter ( a ring) in the disk, we obtain 

$$ d \, \vec{ T  } = {\vec{{\Omega}_{p}}}  \, \wedge \, d \, \vec{L} $$

$$ = {\vec{{\Omega}_{p}}} \, \wedge \, \vec{{\omega}_{1}} \, d \, {I_{1}} + \vec{\Omega_{p}}  \wedge \vec{{\omega}_{2}}  d  {I_{2}} \eqno(16) $$

or

$$ |{d \, \vec{ T  }}_{\perp}|= {\Omega}_{p} \, {\Omega}_{d} \, sin{\delta} \, d \, I_{1} + {{\Omega}_{p}}^2  \, sin{\delta} \, cos{\delta} \,{\left( d \, I_{1} - d \, I_{2} \right)} ,  \eqno(17)$$

with $ {d \, \vec{ \tau }}_{\perp} $ being the differential element of torque perpendicular to the angular momentum,
$ d \, I_{1}$ and $ d \, I_{2}$ , the moments of inertia of the ring.
	
	To obtain the precession velocity for a ring, it suffices to solve only for an element of torque perpendicular to z, say $ {\tau}_{x} $. Clearly,

$$ d \, I_{1}= \rho \, r^2 \, d \, V $$

$$ d \, I_{2}= \rho \, {\left( r^2 \, sin^2 \phi + z^2 \right)} \, d \, V .$$

$$ d \, {\tau}_{x}= \rho \, {\left( - z \, {\left( f_{\phi} \, cos{\phi} + f_{r} \, sin{\phi} \right)} + r \, f_{z} \, sin{\phi} \right)} \, d \, V.  \eqno(18)$$

	Finally, we may write

$$ sin{\delta} \, cos{\delta} \, {\left( r^2 \, cos^2{\phi} - z^2 \right)} \, {{\Omega}_{p}}^2 + r^2 \, {\Omega}_{p} \, {\Omega}_{d} \, sin{\delta} + 
 z \, {\left( f_{\phi} \, cos{\phi} + f_{r} \, sin{\phi} \right)} - r \, f_{z} \, sin{\phi} = 0.  \eqno(19) $$

	A rapid inspection in this equation tell us that:

	a- there is a constraint on$ (r, \phi, z) $ that should be satisfied in order to have a solution for the precession velocity. If this constraint is 
satisfied we may assume the angular precession velocity  approximately constant in time, as long as the time associated to its derivative 
is much longer than the precession period. If these conditions are not met, we should solve the differential Euler equations;
	
	b- if these constraints are satisfied, the precessing disk will admit double solutions. These solutions will be functions  of $ (r, \phi, z, t, \omega) $.

	It should be remarked that this treatment differs, e.g., from that of \citet{ka82}, \citet{pat95} in the sense that
	
	- it takes into account the precessing angular velocity contribution to the angular momentum of the disk,
	- it obtains  constraints, conditions to the validity  of such a treatment, 
	- and the solution is double valued.

\section{The Constraints On The Disk }

	From equation ( 19 ) , of the previous section, one can easily see that the constraints on the precessing disk comes from the inequality

$$ r^4 \, sin ^2{\delta} \, {{\Omega}_{d}}^2 +  4 \, sin{\delta} \, cos{\delta}{\left( r^2 \, cos^2{\phi} \, - z^2 \right)} \, {\tau}_{x} \geq 0. \eqno(20) $$

	Setting $ z= \ell $, where $ \ell$, the height scale of the disk , is related to $ {\rho}$, the density, through

$$ \rho = {\frac{3 \, {\dot M} \, S}{8 \, \pi \, {{\Omega}_{k}} \, {{\ell}^3}}},  \eqno(21) $$

with ${\Omega}_{k}$ being the keplerian velocity, ${\dot M}$ is the accretion rate, S a function that takes into account boundary conditions on the angular momentum and on the 
torque, and is given by

$$ S = 1- {\beta }_{0} \, {\left( \frac{r_{1}}{r} \right)}^{1/2}, $$

$\beta$ being the angular momentum in units of the keplerian one, ${\beta}_{0}$ being this ratio at the inner radius ${\left( 0 { \leq {\beta}} {{\leq 1}}\right)}$,
the inequality may be worked out to read, approximately,

$$
{\beta}^2 \, { r \, sin{\delta} \, M_{p} \, d^2 \, {\rho} - 4 \, r \, K^{2/3} \, sin{\delta} \, {\left( M_{p} + M_{s}\right)} \, {\left( cos{\delta} \, {\left(1+ 3 \, cos{\delta} \, {r\over d}
+ 3 \, sin{\delta} \, {\left( K \over {\rho} \right)}^{1/3} \right)}-1\right)} \, {\rho}^{1/3}}  $$

$$
- 12  K  cos^3{\delta} {r \over d} \geq 0 ,  \eqno(22)
$$

and

$$
{\beta}^2 \,  r \, d^2 \, M_{p} \, sin{\delta} \, {\rho} - 4 \, cos^2{\delta} \, r^2 \, {\left( M_{p}+M_{s} \right)} \, K^{1/3} \, {\rho}^{2/3} + 4 \, cos^2{\delta} \,  {\left( M_{p}+M_{s} \right)} \, K \geq 0, \eqno(23) 
$$ 

where 

$$ K= 7.5 \, 10^{29} \,  {\frac{ {{\dot M}_{1}} \, M_{p} \, r^{3/2} \, S}{\alpha }} \, g . \eqno(24)$$

	The solution to the first inequality is, approximately,

$$ {\rho } \geq 12 cos{\delta} cot{\delta} {\left({{M_{p} + M_{s}}\over {M_{p}}}\right)} {\frac{K}{{\beta}^2 \, {d}^3}}. \eqno(25) $$

	Concerning the second inequality, if

$${\frac{{\left( 16\right)}^2}{27}} \, {\left( {\frac{r}{{\beta} \, d}} \right)}^4 \, cos^2 {\delta} \, cot^2 {\delta} \, {\left({ {M_{p} + M_{s}}\over {M_{p}}}\right)}^2 \leq{1},  \eqno(26)$$

it will hold everywhere in the precessing disk. Therefore, using the definition of $\rho $, we may rewrite 

$$ {\beta}^2 \geq 12 cos{\delta} \, cot{\delta} \, {\left( {M_{p} + M_{s}}\over {M_{p}}\right)} \, {\left( {\ell} \over d \right)}^3  \eqno(27)$$

and 

$$ {\beta}^2 \geq{ {\frac{16}{3 \, \sqrt{3}}} cos{\delta} \, cot{\delta} \, {\left( {{M_{p} + M_{s}} \over {M_{p}}}  \right)}  {\left( {r \over d} \right)}^2} .  \eqno(28)$$ 

	Since $ r \geq {{\ell}} $,  the solution is given by the last expression. Otherwise, if 

$${\frac{{\left( 16\right)}^2}{27}} \, {\left( {\frac{r}{{\beta} \, d}} \right)}^4 \, cos^2 {\delta} \, cot^2 {\delta} \, {\left( {M_{p} + M_{s}}\over {M_{p}}\right)}^2 \geq{1},  \eqno(29)$$

we obtain

$$ 12 cos{\delta} cot{\delta} {\left( {M_{p} + M_{s}}\over {M_{p}}\right)} {\frac{K}{{\beta}^2 \, {d}^3}} \leq {\rho} \leq {K \over r^3} \eqno(30)$$

or 

$$ {\rho} \geq 512 \, cos^3{\delta} \, cot^3{\delta} \, {\left( {\frac{r}{{\beta} \, d}} \right)}^6 \,  {\left( {M_{p} + M_{s}}\over {M_{p}}\right)}^3 \,{ K \over {r}^3}, \eqno(31)$$
which may be rewritten

$$ 12 cos{\delta} \, cot{\delta} \, {\left( {M_{p} + M_{s}} \over {M_{p}}\right)} \, {\left( {r \over d} \right)}^3  \leq {\beta}^2 \leq {\frac{16}{3 \, \sqrt{3}}} cos{\delta} \, cot{\delta} \, 
{\left( {M_{p} + M_{s}}\over {M_{p}}\right)} \, {\left( r \over d \right)}^2 . \eqno(32) $$

	It should be remarked that we may change inequality ( 32 ) into a condition on the masses, which reads

$$ 1.25 {\beta}^2 \, {d_{15}}^2 \leq M_{p} \, {\left( M_{p} + M_{s} \right)} \, \cos{\delta} \, \cot{\delta} \leq{ 0.11 \, {{\beta}^2 \, {d_{15}}^3} \over {{{z}_{d}}^3 \, M_{p}}},  \eqno(33)$$

where $z_{d}$ is the size of the disk in units of Schwarzschild radius and the distance between the primary and the secondary is expressed in units of $ 10^{15}$ cm.

\section{The Average Effect of the Density}

	Our main interest are systems with $r << d$, a situation in which we may treat the tidal torque as a perturbation. Under this condition we may expand 
${A(r, z, t, {\phi}, d, {\omega}, {\delta})}^{-3/2}$, given by (eq.(6), to second order in $r/d$ and $z/d$, and since the constraints obtained in the previous section are satisfied, we may take the average over
$\phi$ and t on eq.(21) .  After some straightforward but tedious algebraic manipulations, we  obtain

$$\sin{\delta} \, \cos{\delta} \, {\left( {r^2\over 2}-z^2 \right)} \, {{\Omega}_{p}}^2+r^2 \, \sin{\delta} \, \beta \, {\Omega}_{k} \, {\Omega}_{p}+
 {{\omega}_{s}}^2 \, \cos{\delta} \, {\left( z^2-r \, {d\over 4} \right)} = 0,  \eqno(34) $$

where ${\omega}_{s}$ is the keplerian velocity of the secondary.

	We now specialize this equation for the average on z and write it in terms of the density $\rho$, i.e.,

$$
{\rho}^{2/3} \, r^2 \, {\left( 6 \, cos{\delta} \, {{\Omega}_{p}}^2+12 \, \beta \, {\Omega}_{k} \, {\Omega}_{p}-3 \,{{\omega}_{s}}^2 \, cot{\delta} \, {d \over r} \right)} = 
2 \, K^{2/3} \, {\left(  2 \, cos{\delta} \, {{\Omega}_{p}}^2-3 \, {{\omega}_{s}}^2 \, cot{\delta} \right)}. \eqno(35)
 $$

To see the behavior of ${\Omega}_{p} $ in terms of the density, let us first  define the parameter p as

$$p={\frac{{\left( 16\right)}^2}{27}} \, {\left( {\frac{r}{{\beta} \, d}} \right)}^4 \, cos^2 {\delta} \, cot^2 {\delta} \, {\left( {M_{p} + M_{s}}\over {M_{p}}\right)}^2 , \eqno(36)$$
 
and to have a better insight into the precession problem in accretion disks and to summarize the results we have obtained so far, we write the 
characteristic densities ${\rho}_{c}, {\rho}_{min}, {\rho}_{*} $ and the actual density, ${\rho}$, in terms of p and $ p_{0}$, the value of p for disks, i.e.,

$${\rho}_{c}={\frac{K}{r^3}} ,$$

$${\rho}_{min}= 3.9 \, {r \over d} \, p^{1/2} \, {\rho}_{c} , $$

$${\rho}_{*}= 17.54 \, p^{3/2} \, {\rho}_{c} ,$$

$${\rho}= {\left( r \over {\ell} \right)}^3 \, {\rho}_{c}= p_{0} \, {\rho}_{c}. \eqno(37) $$

	For $p \rightarrow 0$, there is no constraint and our treatment will be valid for any density. As p grows, but still $0 < p< 1$, there is a characteristic 
density in the problem  and our treatment only holds for ${\rho} \geq {\rho}_{min}$. For $p \geq 1$, there appear two more characteristic densities in the problem, ${\rho}_{c}$ 
and ${\rho}_{*}$.  In the region $ {\rho}_{c} < {\rho} < {\rho}_{*}  $ our treatment doesn't apply as well.

	For disks we are interested in, such as the $ {\alpha}$-standard , slim and advective models, $1 \leq p_{0} < \infty$. The equality at the left occurs when 
 the disk gets thick either due to a huge accretion rate or advection. In that situation, $p_{0} \rightarrow 1$

Now, we plot ${\Omega}_{p}= {\Omega}_{p}(p_{0}) $, assuming $M_{p} >> M_{s}$, ${\delta} \approx 4\raisebox{1ex}{\scriptsize o}$, and $d/r =100$.

Figure 2 above shows that, in the upper branch, $ \Omega_{p} $ decreases in the range $0 <p_{0}< 0.1$, and then increases in $0.1<p_{0}<\infty$;
in the other two branches, ${\Omega}_{p}$ grows with increasing density. For $p_{0} \rightarrow 
 12 cos{\delta} cot{\delta} {\left( {M_{p} + M_{s}}\over {M_{p}}\right)} {\frac{{\ell}^3}{{\beta}^2 \, {d}^3}}, {\Omega}_{p}  \rightarrow \approx  \pm {\omega}_{s} \, {\left( {3 \over 2} \, 
{1 \over sin{\delta}} \right)}^{1/2} $, respectively, in the upper and middle branches. For $ p_{0} \rightarrow {2\over 3} $ , ${\Omega}_{p} \rightarrow \pm  \infty $ 
in the lower and upper branches. In the middle branch, ${\Omega}_{p} \rightarrow  {3\over 8} \, {\frac{{{\omega}_{s}}^2 \, cot{\delta}}{{\beta} }} \, 
{\left({{2 \, d} \over {3 \, r}} -2\right)} $. Asymptotically, 
$ {\Omega}_{p} \rightarrow - {{\beta} \over cos{\delta}} \, {\left( 1\pm {\left(1+{cos{\delta} \over {2 \, {\beta}^2}} \, {{\omega}_{s}}^2 \, cot{\delta} \, {d \over r} \right)}^{1/2} \right)} $.
Angular velocities are in units of the keplerian one.

	Therefore, we will have, in the middle branch,

$$  - {\omega}_{s} \, {\left( {3 \over 2} \, {1 \over {sin{\delta}}} \right)}^{1/2}  \leq {\Omega}_{p} \leq 
{1\over 4} \, {\frac{{{\omega}_{s}}^2 \, cot{\delta}}{{\beta}} \, {d \over r}} , \eqno(38)$$
 in the lower branch,

$$  - \infty < {\Omega}_{p} \leq - {{\beta} \over cos{\delta}} \, {\left( 1+ {\left(1+{cos{\delta} \over {2 \, {\beta}^2}} \, {{\omega}_{s}}^2 \, cot{\delta} \,
 {d \over r} \right)}^{1/2} \right)},  \eqno(39)$$

and, in the upper branch,

$$ +{\omega}_{s} \, {\left( {3 \over 2} \, {1 \over {sin{\delta}}} \right)}^{1/2} \leq {\Omega}_{p} < \infty .  \eqno(40) $$

From equations (38) to (40), we can see a huge spread in the value of the precession velocity due to the effect of the density.

Since for actual disks${\rho}= p_{0} \, {\rho}_{c} $  and $ 0.01 \leq {{\ell} \over r} \leq 1 $, the solutions are at the right of $p_{0}=1$. 
As a matter of fact, for the ${\alpha} $ standard model, the actual density is much greater than ${\rho}_{c} $, the asymptotical behavior is a good approximation.
 Using this limit, we may see that differential precession is drastically reduced as compared to Katz 1973 result.

	It should be remarked that the constraints we have are much more restrictive than those obtained directly from equation (35), which are ,
 approximately,

$${\rho} > 0.356 \, p^{3/4} \, {\rho}_{c} . \eqno(41)$$

	Now, we ask ourselves if, within this formalism, it is possible to recover \citet{ka82} result for the precession velocity, i.e.,

$${\Omega}_{p}^K= -{3 \over 4} \, {{{\omega}_{s}}^2  \over {{\Omega}_{k}}} \, cos{\delta} . \eqno(42)$$

	To answer this question we insert the above expression for the precession velocity into equation (35), to obtain

$$ p_{0} \, {\left( {27 \over 8} \, {\left({\omega}_{s} \over {\Omega}_{k} \right)}^2 \, cos^3{\delta}-9 \, {\beta} \, cos{\delta}-3 \, cot{\delta} \, {d \over r} \right)} =
{9 \over 4} \, {\left({\omega}_{s} \over {\Omega}_{k} \right)}^2 \, cos^3{\delta}-6 \, cot{\delta} . \eqno(43)$$

	Since ${\omega}_{s} << {\Omega}_{k} $, it implies $p_{0} \approx {{2 \, r} \over d} $, or 

$${\rho}^{k} \approx 2 \, \sqrt{2} \, {\left( r \over d \right)}^{1.5} \, {\rho}_{c} . \eqno(44) $$ 

 Therefore, we must conclude that, under our formalism, \citet{ka82} results only applies to disks with density profile quite different from that expected 
under the keplerian ${\alpha} $ thin disk approximation. As a matter of fact, the results are in complete disagreement with this approximation.

\section{The Importance of deviations from the mean precessional motion}

	Differently from \citet{ka82}, in ou formalism, deviations from the mean precessional motion can't be treated as perturbations, i.e., nodding motions. 
As will be shown, the amplitude of these motions, though small for the precessional time scale, is quite large, as far as the appropriate time scale is considered. 
As a matter of fact, these deviations may not be periodic motions, but motions in which misalignment increases with time.  The prevailing scenario will 
depend on the time scales involved.

	The importance of the misalignment motion  in our formulation can be seen in the time evolution of $\delta$. From equation (15),

$${\dot {\delta}}= {\Omega}_{p} \, sin{\delta}. \eqno(45)$$

	Now, suppose that for some unespecified reason, ${\Omega}_{p} $ is given by the extreme right of the middle branch.  Then, the time evolution will 
be given by the solution of 

$${\dot {\delta}}= {1\over 4} \, {{{\omega}_{s}}^2 \over {{\beta} \, {\Omega}_{k}}} \, cos{\delta} \, {d \over r} \, ,  \eqno(46)$$

which is

$$ sin{\delta}=  {{ B \, exp {\left ( {1\over 2} \, {{{\omega}_{s}}^2 \over {{\beta} \, {\Omega}_{k}}} \,  {d \over r} \, t \right)}-1} \over {
B \, exp {\left ( {1\over 2} \, {{{\omega}_{s}}^2 \over {{\beta} \, {\Omega}_{k}}} \,  {d \over r} \, t \right)}+1}} , \eqno(47)$$

where B is an integration constant. From Figure 1, we see that ${\delta}$ is the angle between z and z', respectively the normal to the plane of the disk and the 
normal to the plane of the secondary orbit. ${\omega }_{2}$ lies in the plane zz', and is normal to z. Therefore, if ${\delta}_{0}$  is the angle between z and z'at $t=0$, 
we must have, at any time,

$$cos {\delta}= cos{\delta}_{0} \, cos{\Theta}_{2}, \eqno(48)$$

where ${\Theta}_{2} $ is the angle in a plane that contains z and is normal to ${\omega}_{2}$ . Assuming that for $t \rightarrow 0$, ${\Theta}_{2} \rightarrow 0$,
 we obtain 

$$B= {\frac{1}{cos^2{\delta}_{0}}} \, {\left( 1+sin^2{\delta}_{0} \pm 2 \, sin{\delta}_{0} \right)} , \eqno(49)$$

and 

$$cos{{\Theta}_{2}}= {\frac{2}{cos{\delta}_{0}}} \, {\left( \frac{B^{1/2} \, {\exp b \, t}}{B \, {\exp{2 \, b \, t}}+1} \right)} ,\eqno(50)$$

where

$$b={1\over 2} \, {{{\omega}_{s}}^2 \over {{\beta} \, {\Omega}_{k}}} \,  {d \over r}.$$

 We see that, as time goes on, the misalignment  increases.

	Now, if we take our solution from the lower branch, assuming the unperturbed disk to be quasi keplerian,

$${\dot {\delta}}= - 2 \, {\beta} \, {\Omega}_{k} \, tan {\delta},  \eqno(51)$$

and 

$$ sin {\delta}= B \, exp  2 \, {\beta} \, {\Omega}_{k} \, t .  \eqno(52)$$

Proceeding in the same way as we did before, imposing the same boundary conditions, we obtain $B^2= sin^2{\delta}_{0}$ and

$$cos^2{\Theta}_{2}= \frac{1}{cos^2{\delta}_{0}} \, {\left( 1-sin^2{\delta}_{0} \, {\exp {2 \, {\beta} \, {\Omega}_{k} \, t}} \right)}. \eqno(53)$$

It is easily seen that this solution holds for $t < t_{0}$, where 

$$t_{0}= - \frac{1}{ {\beta} \, {\Omega}_{k}} \, \ln{|sin{\delta}_{0}|},\eqno(54)$$

and for $ {\beta} \approx 1$, in a very fast way, ${\Theta}_{2} \rightarrow 0.5 \, \pi  $.

	Finally, let us analyse the situation $\beta$ very small in the middle branch. This implies

$$ {\Omega}_{p} \approx  {\omega}_{s} \, {\left( \frac{d}{2 \, r \, sin{\delta}} \right)}^{1/2}, \eqno(55) $$

and 

$${\dot {\delta}}=A \, {\left( sin{\delta} \right)}^{1/2}, \eqno(56)$$

where A is defined in the previous equation. Making the substitution
$$cos{\delta}_{0} \, cos{\Theta}_{2}= x, \eqno(57) $$

yields

$$\frac{\dot x}{{\left( 1-x^2 \right)}^{3/4}} =A. \eqno(58)$$

	Since we are only interested in the behavior of the solution,  we make the approximation

$$<{\left( 1-x^2 \right)}>^{1/4} \frac{\dot x}{{\left( 1-x^2 \right)}} \approx A,\eqno(59)$$

where $<   >$ stands for an average. Then,

$$ cos{\Theta}_{2}= \frac{1}{cos{\delta}_{0}} \, {\left( \frac{C \, \exp{-2 \, b \, t}-1}{C \, \exp{-2 \, b \, t} +1} \right)}, $$

$$C=\frac{1+cos{\delta}_{0}}{1-cos{\delta}_{0}}, $$

$$b=\frac{A}{<{\left(1-x^2 \right)}>^{1/4}}. \eqno(60)$$

	Again, the solution only holds for $t < t_{0}$, where

$$t_{0}= \frac{1}{2 \, b} \, \ln {\left( \frac{1+cos{\delta}_{0}}{1-cos{\delta}_{0}} \right)}. \eqno(61)$$

	To obtain the time evolution of $\delta$ we have assumed that the time scales associated to it are much longer than the remaining time scales in the 
problem, so any function of $\delta$ survives to the averaging process. In other words, even after the averaging process, for any$ f=f(\delta)$,  
$\delta$ is considered the instantaneous value. However, if the time scales are comparable, equation (45) should be interpreted as

$${\dot {\delta}}= <{\Omega}_{p} \, sin{\delta}>, \eqno(62)$$

the solution being

$${\delta}= <{\Omega}_{p} \, sin{\delta}> \, t + {\delta}_{0}, \eqno(63)$$

and 

$$cos{\Theta}_{2}= cos {\left( <{\Omega}_{p} \, sin{\delta}> \, t \right)}-tg{\delta}_{0} \, sin{\left( <{\Omega}_{p} \, sin{\delta}> \, t \right)}. \eqno(64)$$

	Clearly, we have not considered viscosity in our formalism. Its inclusion certainly invalidates the use of eq.(45) to study the temporal evolution of the 
 inclination angle ${\delta}$. We should use, instead, a second order differential equation.

\section{How  Keplerian Thin Disks Precess?}

	Most of the work on the literature assume an keplerian disk precessing under the perturbing influence of a third body, and to answer that question 
we shall take a naive approach to the angular momentum transport in the disk.
	If $ {\dot j}_{in} $, $ {\dot j}_{0} $ and ${\dot j}_{out} $ are, respectively, the inwards angular momentum transport ( per unit time), the rate at which the angular momentum flows 
into the central compact object and the outwards flux of angular momentum, we have

$${\dot j}_{in} = {\dot j}_{0}+{\dot j}_{out} ,  \eqno(65)$$
 and we may write

$$ {\dot j}_{in} = {\beta} \, {\dot M} \, \sqrt { G \, M \, r } , \eqno(66)$$

$$ {\dot j}_{0} = {\beta}_{0} \, {\dot M} \, \sqrt { G \, M \, r_{1} } \eqno(67) $$
and 

$$ {\dot j}_{out} = 2 \, \pi \, r \, {\ell} \, {\tau}_{r \, \phi} ,  \eqno(68)$$

where  $ {\tau}_{r \, \phi}$ is the stress tensor given by

$${\tau}_{r \, \phi} = -2 \, \eta \, {\sigma}_{r \,, \phi}  ,  \eqno(69)$$

$ {\sigma}_{r \,, \phi} $ being the rate of strain tensor, given by

$${\sigma}_{r \,, \phi} = - {r \over 2} \, {\partial {{\beta} \, V_{k}} \over {\partial r}}, \eqno(70)$$

where $V_{k} $ is the keplerian velocity.

	Then, the angular momentum conservation may be written as

$$ {\beta} \, {\dot M} \, \sqrt { G \, M \, r }= {\beta}_{0} \, {\dot M} \, \sqrt { G \, M \, r_{1} }+2 \, \pi \, r \, {\ell} \, {\tau}_{r \, \phi}.  \eqno(71)$$

In the above equation, we have assumed null boundary condition for the torque at the inner radius $r= r_{1}$. This yields for the energy dissipation, neglecting 
$ {\beta} $ dependence on r,

$$ q^{+} = {3 \over {8 \, {\pi}}} \, {{G \, M \, {\dot M}} \over r^3} \, {\beta}^2 \, {\left(1-{{\beta}_{0} \over {\beta}} \, {\left( {r_{1} \over r} \right)}^{1/2} \right)} , \eqno(72)$$

and for the density

$$\rho = {3 \over {8 \, {\pi}}} \, {{{\dot M} \, {\beta}} \over {{\Omega}_{k} \, {\ell}^3}} \,  {\left(1-{{\beta}_{0} \over {\beta}} \, {\left( {r_{1} \over r} \right)}^{1/2} \right)} . \eqno(73)$$

	Now, we write the energy conservation, as matter in the disk moves from a point $r_{2} $ to a point r, as 

$$ {\Delta E}_{p} +{\Delta E}_{k} +{\Delta E}_{int}+Q^{-} =0,  \eqno(74)$$

where ${\Delta E}_{p} $ is the variation of potential energy,  ${\Delta E}_{k} $ the variation of kinetic energy, ${\Delta E}_{int}$ is the  variation of the 
internal energy and $ Q^{-}$ is the total amount of energy that leaves the system  from $ r_{2} $ to r. 
	Writing for the radial velocity
$$V_{r}= -{\alpha} \, {\left( {{\ell} \over r} \right)}^2 \, V_{k} ,  \eqno(75)$$

we obtain, assuming $r_{2} >> r$, and that the heat produced by shear leaves the system,

$$ -{\dot M} \, {{G \, M} \over r}+ {{\dot M} \, {V_{k}}^2 \over 2} \, {\left( {\beta}^2+{\left( {\ell} \over r \right)}^2 +{\alpha}^2 \, {\left( {\ell} \over r \right)}^4 \right)}
+{3 \over 2} \, {\beta}^2 \, {\dot M} \, {{G \, M} \over r}{\left(1-{{\beta}_{0} \over {\beta}} \, {\left( {r_{1} \over r} \right)}^{1/2} \right)}=0 ,  \eqno(76)$$

where, again, we have neglected ${\beta} $ dependence on r . Collecting like terms, we get

$$ 4 \, {\beta}^2-2 \, {\beta}_{0} \, {\beta} \, {\left( r_{1} \over r \right)}^{1/2}+{\left( {\ell} \over r \right)}^2+{\alpha}^2 \, {\left( {\ell} \over r \right)}^4-2 =0, \eqno(77) $$

which tells us that the keplerian approximation only holds for $r=1$ and $ {\alpha}^2 \, {\left( {\ell} \over r \right)}^4 <<1 $. Under the assumption of thin 
disk and $\alpha << 1$, for $r \rightarrow \infty $, $ {\beta} \rightarrow \approx 0.7$.

	Now, assuming $ {\beta} \approx constant $, and  using inequality (29), i.e., 

$$ {\beta}^2  \leq {\frac{16}{3 \, \sqrt{3}}} cos{\delta} \, cot{\delta} \, {\left( {M_{p} + M_{s}}\over {M_{p}} \right)} \, {\left( {r \over d} \right)}^2 , \eqno(78)$$        

leads to 

$$ {\left( d \over r \right)}^2 \leq {64 \over {3 \, \sqrt{ 3 }}} \, {{cos{\delta} \, cot{\delta}} \over {2-{\left( {\ell} \over r \right)}^2-{\alpha}^2 \, {\left( {{\ell} \over r}  \right)}^4}} \, {\left( {M_{p} + M_{s}}\over {M_{p}} \right)}  , \eqno(79)$$

and, since $ d >> r $ , this implies 
$$ {\alpha}^2 \, {\left( {\ell} \over r \right)}^4+{\left( {\ell} \over r \right)}^2 \approx 2 ,$$
 or, approximately, since $ \alpha << 1 $ ,  $ {\ell} \approx r,$  which means the breakdown of the keplerian and thin disk approximations. Despite not having 
treated rigorously the angular momentum transport, this result is quite general, as it may be seen by considering the r-component of the force equation, 

$${1/2} \,{{ {\partial} \,{ V_{r}}^2} \over {{\partial} \,r}}-{\beta}^2 \,{{V_{k}}^2 \over r }=-{1 \over {\rho}} \, {{\partial P} \over {\partial r}}-{{G \, M} \over r^2}, \eqno(80)$$

which, after using the expression for ${\rho}$ when the disk gets thick, is integrated to give

$$3 \, {\alpha}^2 \, {\left( {\ell} \over r \right)}^4+5 \, {\left( {\ell} \over r \right)}^2-6 \, {\left(1-{\beta}^2 \right)}=0, \eqno(81)$$

yielding

$${\left( {\ell} \over r\right)}^2 \approx {6 \over 5} \, {\left(1-{\beta}^2 \right)}. \eqno(82)$$

	To analyse the situation $ p <1 $, the sign of inequality (78) is inverted, i.e.,

$$ {\left( d \over r \right)}^2 \geq {64 \over {3 \, \sqrt{ 3 }}} \, {{cos{\delta} \, cot{\delta}} \over {2-{\left( {\ell} \over r \right)}^2-{\alpha}^2 \, {\left( {{\ell} \over r}  \right)}^4}} \, {\left( {M_{p} + M_{s}}\over {M_{p}} \right)}  , \eqno(83)$$

and since $d/r>>1$, a trivial solution is ${\ell}/r << 1$.  Therefore, we are led to the conclusion that $p>1$ implies ${\beta}<<1$, and $p<1$, ${\beta} \approx 1$.

 it should be reminded  that the inner region of accretion disks around supermassive black holes gets thicker either due to the radiation pressure
 dominance or to the ineffective radiative cooling, a situation in which advection becomes the dominant cooling mechanism, and $ {\beta} <1$.

	From these results, and those of the previous section, we must conclude that advective disks will precess with a velocity close to that given by 
$\rho \approx {\rho}_{c} $, i.e.,

$$ {\Omega}_{p} \approx  -{3  \over cos{\delta}}\, {\left( {\beta} -{\left({\beta}^2+{1/6} \, {{\omega}_{s}}^2 \, cot{\delta} \, {\left({{2 \, d} \over {3 \,
 r}} -2\right)} \right)}^{1/2} \right)}  , \eqno(84) $$

and the standard-$\alpha$ keplerian thin disk will precess, roughly, with the asymptotic value of the precessing velocity, i.e.,

$$ {\Omega}_{p} \approx  {1\over 4} \, {\frac{{{\omega}_{s}}^2 \, cot{\delta}}{{\beta} }} \, {\left( d \over  r \right)}  . \eqno(85) $$

\section{The p- profile of the accretion disk or precession and disk model}

	The rate of precession depends on the kind of disk model we are assuming. Different models give different relations between p and $p_{0}$, as a
 consequence different precession rates. Besides, we have seen $\beta$ is related to the scale height of the disk through 
$${\beta}^2= 1-{\left({\ell} \over r \right)}^2,$$

which means

$${\beta}^2= 1-{p_{0}}^{-2 / 3}. \eqno(86)$$

	Inserting this relation into the definition of p, equation (62), to rewrite it as

$$p=  {\frac{256}{27}} cos^2{\delta} \, cot^2{\delta} \, {\left( {M_{p} + M_{s}}\over {M_{p}} \right)}^2 \, {\left( {r \over d} \right)}^4  \,
{\frac{p_{0}^{4/3}}{{\left({p_{0}}^{2/3}-1 \right)}^2}}. \eqno(87)$$

	For the ${\alpha}$-standard model, assuming black-body emission and gas pressure dominance

$$p_{0} \approx 4.53 \times 10^{11} \, {\left( M_{9} \over {{\dot M}_{1} \, z} \right)}^{3/8} >> 1, \eqno(88)$$

and 

$$p=  {\frac{256}{27}} cos^2{\delta} \, cot^2{\delta} \, {\left( {M_{p} + M_{s}}\over {M_{p}} \right)}^2 \, {\left( {r \over d} \right)}^4  <<1. \eqno(89)$$

$M_{9}$ and ${\dot M}_{1}$ are, respectively, the mass of the primary in units of $10^9 $ solar masses and the accretion rate in units of 1 solar mass per year. 

Therefore, under these assumptions, the disk is keplerian and our formalism holds everywhere. The precession velocity will be given by the asymptotic 
limit of the middle branch, with ${\beta}=1$,

$${\Omega}_{p} \approx {1 \over 4} \, {{{\omega}_{s}}^2 \over {\Omega_{k}}} \, {d \over r} \, cot {\delta}. \eqno(90) $$

	To finalize, we can easily conclude that the $\alpha$-thin keplerian disk is not affected by the constraints on the density. This, certainly, is not the
 case if the disk is thick.  Let us now examine a disk radiative pressure dominated. Let us also assume cooling due to radiation. In that case,

$$P \approx P_{r}, $$

$$P_{r}= \frac{{\rho} \,{\sigma}_{T} }{c \, m_{H}} \, {\ell} \, F_{r}, $$

$$P= \frac{1}{3} \, {\rho} \, {\Omega_{k}}^2 \, {\ell}^2. \eqno(91)$$

	This implies, using equation (72 ), approximately, 

$$\frac{{\ell}}{r} = \frac{9 \,{\rho} \,{\sigma}_{T} }{8 \, \pi \, c \, m_{H}} \, \frac{{\dot M}}{r} \, {\left(1-{\left( \frac{{\ell}}{r} \right)}^2 \right)} , \eqno(92)$$

the solution being

$$\frac{{\ell}}{r}= \frac{1}{2 \, a} \, {\left( {\left(1+4 \, a^2 \right)}^{1/2}-1 \right)} , \eqno(93)$$

or

$${p_{0}}^{1/3}= { \frac{2 \, a}{{\left( {\left(1+4 \, a^2 \right)}^{1/2}-1 \right)}}} \,  , \eqno(94)$$

where 

$$a= {\frac{0.34}{z}} \, \frac{{\dot M}_{1}}{M_{9}}, \eqno(95)$$

 and  z, now, is the radial distance in units of $3 \, R_{s}$.

	In figure 3, below, we plot $p_{0}=p_{0}(a) $.

	Using equation (87 ), and imposing ${\rho}= {\rho}_{*}$, we obtain

$${p_{0}}^{4/3}-2 \, {p_{0}}^{2/3}+1=414.45 \, cos^2{\delta} \, cot^2{\delta} \, {\left( \frac{M_{p}+M_{s}}{M_{p}} \right)}^2 \, {\left( r \over {d} \right)}^4 \, {p_{0}}^{4/3} , \eqno(96)$$

which, using equation (94 ) above, changes to

$$ {\left( {\left(1+4 \, a^2\right)}^{1/2}-1 \right)}^4-8 \, a^2 \, {\left( {\left(1+4 \, a^2\right)}^{1/2}-1 \right)}^2+16 \, a^4= $$

$$ 6631.2 \, cos^2{\delta} \, cot^2{\delta} \, {\left( \frac{M_{p}+M_{s}}{M_{p}} \right)}^2 \, {\left( z \over {d_{15}} \right)}^4 \, a^4 . \eqno(97)$$

	Expressing $\frac{z}{d_{15}}$ in terms of the precession period, $T_{p}$, and z in terms of a, yields

$${\left( \frac{z}{d_{15}} \right)}^4= {\frac{1.95\times 10^{-5}}{cos^2{\delta} \, {T_{p}}^2}} \, {\frac{{{\dot M}_{1}}^3}{a^3 \, {\left( M_{p}+M_{s} \right)}^2 \, {M_{p}}^3}} \, {\left( \frac{{p_{0}}^{2/3}-1}{{p_{0}}^2/3} \right)}. \eqno(98)$$

Inserting into equation (97 ) and defining

$$y= \frac{ 3.13 \times 10^{-3} \, cos^2{\delta} \, {{\dot M}_{1}}^3}{{T_{p}}^2 \, {M_{9}}^5}, \eqno(99)$$

we get

$$y= a \, {\left( 4 \, a^2-{\left( {\left(1+4 \, a^2 \right)}^{1/2}-1 \right)}^2 \right)} . \eqno(100)$$

	Below, a plot of $y= y(a)$.

	A glance at this figure shows that if we are looking for a solution with $p_{0} \approx 1$, we should choose around $a \approx 25.74$. This gives

$$p_{0} \approx 1.06,$$

$$z_{d} \approx 0.0132 \, \frac{{\dot M}_{1}}{M_{9}}, $$

$$d_{15} \approx 5.07 \,{\left( \frac{{{\dot M}_{1}}^4 \, {\left(M_{p}+M_{s} \right)}^2 \, cos^2{\delta} \, cot^2{\delta}}{{M_{p}}^6} \right)}^{1/4} , $$

$$T_{p}= 2.81\times 10^{-3} \, cos{\delta} \, {\left( \frac{{{\dot M}_{1}}^3}{{M_{9}}^5} \right)}^{1/2} , \eqno(101)$$

where $z_{d}$ is the size of the disk in units of $3 \times R_{s}$.

It is worth remarking that an acceptable solution gives $z > 1$, or $ \frac{{\dot M}_{1}}{M_{9}} >  75.76$ if the primary is a Schwarzschild black hole, or 
 $ \frac{{\dot M}_{1}}{M_{9}} >  25.25$ for an extremely rotating Kerr black hole. Assuming these constraints do apply, we may write

$$p \approx  0.15 \, {\left( z \over z_{d} \right)}^4, 1\leq z \leq z_{d}.\eqno(102) $$

	However, if they don't apply, $\rho > {\rho}_{*}$ everywhere. Then, use of equations (89 ) and( 98 ) will lead to 

$$p =2.99 \, 10^{-3} \, cos^2{\delta} \, \frac{z^3}{{T_{p}}^2 \, {M_{9}}^2} \, \frac{a^2}{{\left(4 \, a^2-{\left( {\left(1+4 \, a^2 \right)}^{1/2}-1\right)}^2 \right)}} .\eqno(103)$$

	Since p given above has to be smaller than p given by equation (102  ), we obtain a lower bound on $T_{p}$, given by

$${T_{p}}^2=0.02 \, cos^2{\delta} \, \frac{{z_{d}}^3}{{M_{9}}^2} \, \frac{a^2}{{\left(4 \, a^2-{\left( {\left(1+4 \, a^2 \right)}^{1/2}-1\right)}^2 \right)}} . \eqno(104)$$

	Below, we plot ${T_{p}}^2$ for $\delta \approx 4$, $z_{d}=100$ and $M_{9}=6.0$.

	Finally, if the disk is radiation pressure dominated and cooled by advection, equation (92 ) changes to

$$a \, {\left( {\ell} \over r \right)}^2+{\left(1-2.2 \, a \right)} \, {\left( {\ell} \over r \right)}-a=0, \eqno(105)$$

the solution being

$$\frac{{\ell}}{r}= \frac{1}{2 \, a} \, {\left( {\left(1-4.4 \, a +8.84 \,a^2 \right)}^{1/2}-{\left(1-2.2 \, a \right)}  \right)} ,\eqno(106)$$

or

$${p_{0}}^{1/3}= \frac{{\left(1-4.4 \, a+8.84 \,  a^2 \right)}^{1/2}+{\left(1-2.2 \, a \right)}}{2 \, a}  , \eqno(107)$$

and equation ( 100 ) changes to

$$y= a^3 \, {\left( {\left( {\left(1-4.4 \, a+8.84 \,  a^2 \right)}^{1/2}+{\left(1-2.2 \, a \right)} \right)}^2-4 \, a^2 \right)} /$$

$${\left( {\left(1-4.4 \, a+8.84 \,  a^2 \right)}^{1/2}+{\left(1-2.2 \, a \right)} \right)}^2, \eqno(108)$$

where y and a have been previously defined.

	Below, we plot $p_{0}=p_{0}(a)$ and $y=y(a)$.

	Proceeding as before, we look for a solution with $p_{0}  \approx 1$. This will force us to look around $a\approx 0.4525$, yielding

$$p_{0}\approx 1.01,$$
$${\beta} \approx 0.1,$$
$$y \approx 5\times 10^{-3}, $$
$$z_{d} \approx 0.75 \frac{{\dot M}_{1}}{M_{9}}, $$
$$T_{p} \approx 0.626 \, cos{\delta} \, {\left( {{\dot M}_{1}}^3 \over {M_{9}}^5 \right)}^{1/2}, $$
$$d_{15} \approx 10.38 \, {\left( cos^2{\delta} \, cot^2{\delta} \, {{\dot M}_{1}}^4 \,  \frac{{\left(M_{9}+M_{s} \right)}^2}{{M_{9}}^6} \right)}^{1/4}. \eqno(109)$$

	Reasoning as we did before, $\frac{{\dot M}_{1}}{M_{9}} >1.33$ if the primary is a Schwarzschild black hole and $\frac{{\dot M}_{1}}{M_{9}} >0.22$ for
 an extremely rotating Kerr blak hole. The p profile in the disk is still given by equation (102). Again, if these constraints on the density don't apply, p will be

$$p =4.7 \, 10^{-3} \, cos^2{\delta} \, \frac{z^3}{{T_{p}}^2 \, {M_{9}}^2} \, {{\left( {\left(1-4.4 \, a+8.84 \, a^2 \right)}^{1/2}+1-2.2 \, a \right)}^2\over
 {\left( {\left( {\left(1-4.4 \, a+8.84 \, a^2 \right)}^{1/2}+1-2.2 \, a \right)}^2-4 \, a^2 \right)}}, \eqno(110)$$

and  the lower bound on the  precession period 
$${T_{p}}^2=0.031 cos^2{\delta} \, \frac{z^3}{{M_{9}}^2} \, {{\left( {\left(1-4.4 \, a+8.84 \, a^2 \right)}^{1/2}+1-2.2 \, a \right)}^2\over
 {\left( {\left( {\left(1-4.4 \, a+8.84 \, a^2 \right)}^{1/2}+1-2.2 \, a \right)}^2-4 \, a^2 \right)}}. \eqno(111)$$

	Figure 5 gives $T_{p}=T_{p}(a)$ for ${\delta} \approx 4$, $z_{d}=100$ and $M_{9}=6.0$.

\section{Conclusions}

	We have treated the problem of precession in accretion disks taking into account its contribution to the total angular momentum. We have looked for 
conditions under which the problem may be treated by solving the algebraic Euler equations, i.e., constant angular velocities. We have found that the problem is 
characterized by the parameter p, given by

$$p=  {\frac{256}{27}} {{cos^2{\delta} \, cot^2{\delta}} \over {\beta}^4} \, {\left( {M_{p} + M_{s}}\over {M_{p}} \right)}^2 \, {\left( {r \over d} \right)}^4 ,$$ 

and the densities

$${\rho}_{min}=  12 cos{\delta} cot{\delta} {\left({{M_{p} + M_{s}}\over {M_{p}}}\right)} {\frac{K}{{\beta}^2 \, {d}^3}}, $$

$${\rho}_{c}= {K \over r^3} ,$$

$${\rho}_{*}=  512 \,  cos^3{\delta}  cot^3{\delta}   {\left( {M_{p} + M_{s}}\over {M_{p}} \right)}^3  {\left( {r \over {{\beta}  d}} \right)}^6 {K \over r^3} .$$

	For $p \leq 1$, the only constraint is $\rho \geq {\rho}_{min} $, and the solutions, as a function of the density,  lie in three branches: the upper one,
 with prograde precession velocities, at the upper left of $\rho ={2 \over 3} {\rho}_{c} $; the lower branch, with retrograde velocities, at the lower right of 
$\rho ={2 \over 3} {\rho}_{c} $; and the middle branch, between the upper and the lower, in which the solution is retrograde for $\rho < {r \over d} {\rho}_{c} $,
 and prograde for $\rho  > {r \over d} {\rho}_{c} $. For $ p \geq 1$, part of the $\rho $-space, $ {\rho}_{c} \leq {\rho} \leq {\rho}_{*} $, is not allowed. Using simple 
energy and angular momentum transport arguments, together with constraints obtained in that formulation, it is shown that, for $p >1$, $ \alpha \rightarrow 1$ and 
 ${\ell} \rightarrow r$, leading to the breakdown of the keplerian and thin disk approximations. In that situation the disk is luminosity defficient.  As a matter of fact, a very important contribution of this paper 
is to show the incompatibility between keplerian disk and thick disk. 
	The procedure we have adopted is suitable for treating deviations from the mean precessional motion, the misalignment motion, as well. For 
different limiting expressions for the precession velocity, we have 
obtained the time evolution for $ \delta$, the angle that measures the misalignement of the plane of the disk and the orbital plane of the secondary star,
 the misalignment  angle. It has been shown that the time evolution of ${\delta}$ depends on the magnitude of the time scale associated to this motion.
Periodic motions only occur if this time scale is comparable to the remaining ones. Otherwise, there will be a tendency to force the plane of the disk to 
become normal to the secondary orbital plane.   It is shown that, if the disk is keplerian, full misalignment, i.e. ${\delta}= 0.5 \, \pi$, is reached in a time 
comparable to the precession period, which means that the solution for the system is no longer given algebraically. We should make resort to the solutions
 of the differential Euler equations, a formidable task far beyond our goal in this paper. 
	Finally, for different disk models we have shown that the constraints on the density will hold as long as ${{\dot M}_{1} \over M_{9}} >> 1$. In that situation, 
we have found the properties of the precessing region as a function of ${\dot M}_{1}$ and $ M_{9}$. An expression for the separation distance between the primary 
and the secondary is also found. Otherwise, if ${{\dot M}_{1} \over M_{9}} < 1$, lower bounds on the precession period and on the separation distance are 
found. These bounds depend on ${\dot M}_{1} \over M_{9}$.  Application of this formalism to different disk models, may be summarized as follows:
 1- for the keplerian thin $\alpha$ standard disk model, the precession period (in years) will be

$$ T_{p} \geq  0.08 \, z_{d}^{1.5}  \,  M_{p}, $$

 For accretion disks in binary systems, we may take the size of the disk as the truncation radius, the point where the Roche equipotentials first intersect. 
In our systems we are not allowed to do so but, in any case, we found in the literature  that values for $ z \approx 100-1000 $ are not unusual. Besides,
 expecting precessiuon periods of about hundred years, the system will reach full misalignment in a time of the order of the precession period.  The disk is 
not affected by constraints on the density.

2- For disk in which radiation dominates pressure and cooling, if ${{\dot M}_{1} \over M_{9}} > 75.76$ for a Schwarzschild black hole and ${{\dot M}_{1} \over M_{9}} > 12.62$ 
 for a Kerr black hole, density in the disk may come close to the critical density ${\rho}_{*}$. In that situation, ${\ell} \approx r$, and the size of the disk can not exceed 

$$z_{d}= 0.0132 \, {{\dot M}_{1} \over M_{9}}.$$

Besides, if the system is not to be affected by gravitational radiation, ${{\dot M}_{1} \over M_{9}}$ should be much larger than the values given above. 
Precession period will be 

 $$T_{p}= 2.81\times 10^{-3} \, cos{\delta} \, {\left( \frac{{{\dot M}_{1}}^3}{{M_{9}}^5} \right)}^{1/2} .$$

A reasonable value for $z_{d}$ would imply a very large ratio ${{\dot M}_{1} \over M_{9}}$. We are, therefore, led to the suspicion that the assumption 
$ {\ell} \approx r$, or $p_{0} \approx 1$, is very strong for this model. Abandon of this assumption leads to a lower bound on the precession period, given by

$${T_{p}}^2=0.02 \, cos^2{\delta} \, \frac{{z_{d}}^3}{{M_{9}}^2} \, \frac{a^2}{{\left(4 \, a^2-{\left( {\left(1+4 \, a^2 \right)}^{1/2}-1\right)}^2 \right)}} .$$

3-Finally, for a radiation pressure and advective cooling dominated disk, $ {\ell} \approx r$, and its size

$$z_{d}= 0.75 \, {{\dot M}_{1} \over M_{9}}.$$

Again, if the system is not to be affected by gravitational radiation, otherwise it will live for a short period of time, ${{\dot M}_{1} \over M_{9}} >> 1$.
If ${{\dot M}_{1} \over M_{9}}$ is not too large, ${\ell} << r$,  and the constraints on the density will not be effective. Again, there will be a lower bound on
the precession period given by

$${T_{p}}^2=0.031 cos^2{\delta} \, \frac{z^3}{{M_{9}}^2} \, {{\left( {\left(1-4.4 \, a+8.84 \, a^2 \right)}^{1/2}+1-2.2 \, a \right)}^2\over
 {\left( {\left( {\left(1-4.4 \, a+8.84 \, a^2 \right)}^{1/2}+1-2.2 \, a \right)}^2-4 \, a^2 \right)}}.$$

	To finalize, we would stress the main contribution of this work as pointing out the possibility of considering the disk structure, when studying 
disk precession, by means of an alternative procedure.




\clearpage                         


\begin{figure}
\plotone{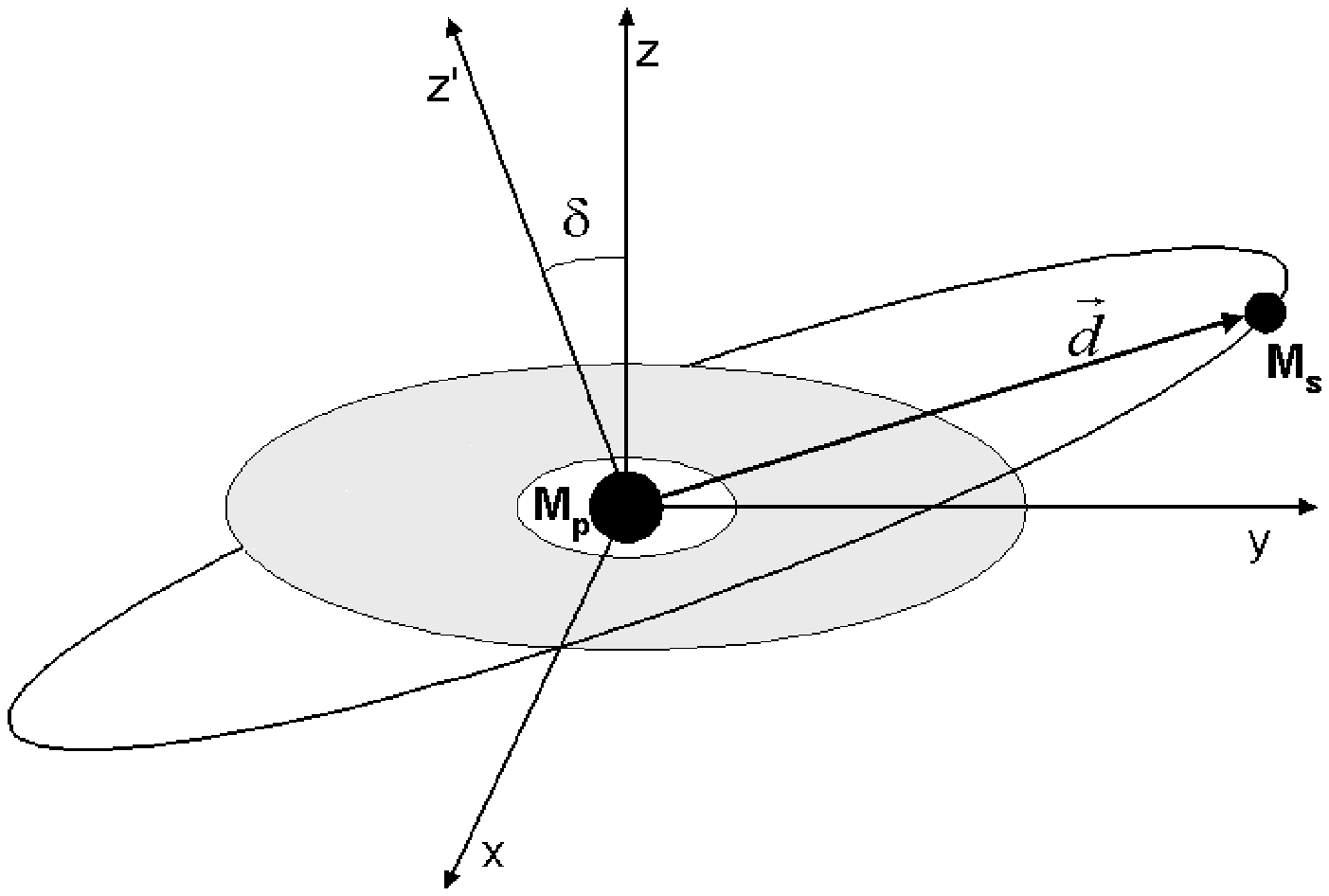}
\caption{Binary System Geometry. P stands for primary and S for secondary; zand z' are, respectively, normal to the plane of the disk and to the secondary
 orbital plane;  ${\delta}$ is the inclination angle between them; d is the secondary distance to the primary. The line of nodes is coincident with x}
\label{fig1}
\end{figure}

\clearpage

\begin{figure}
\plotone{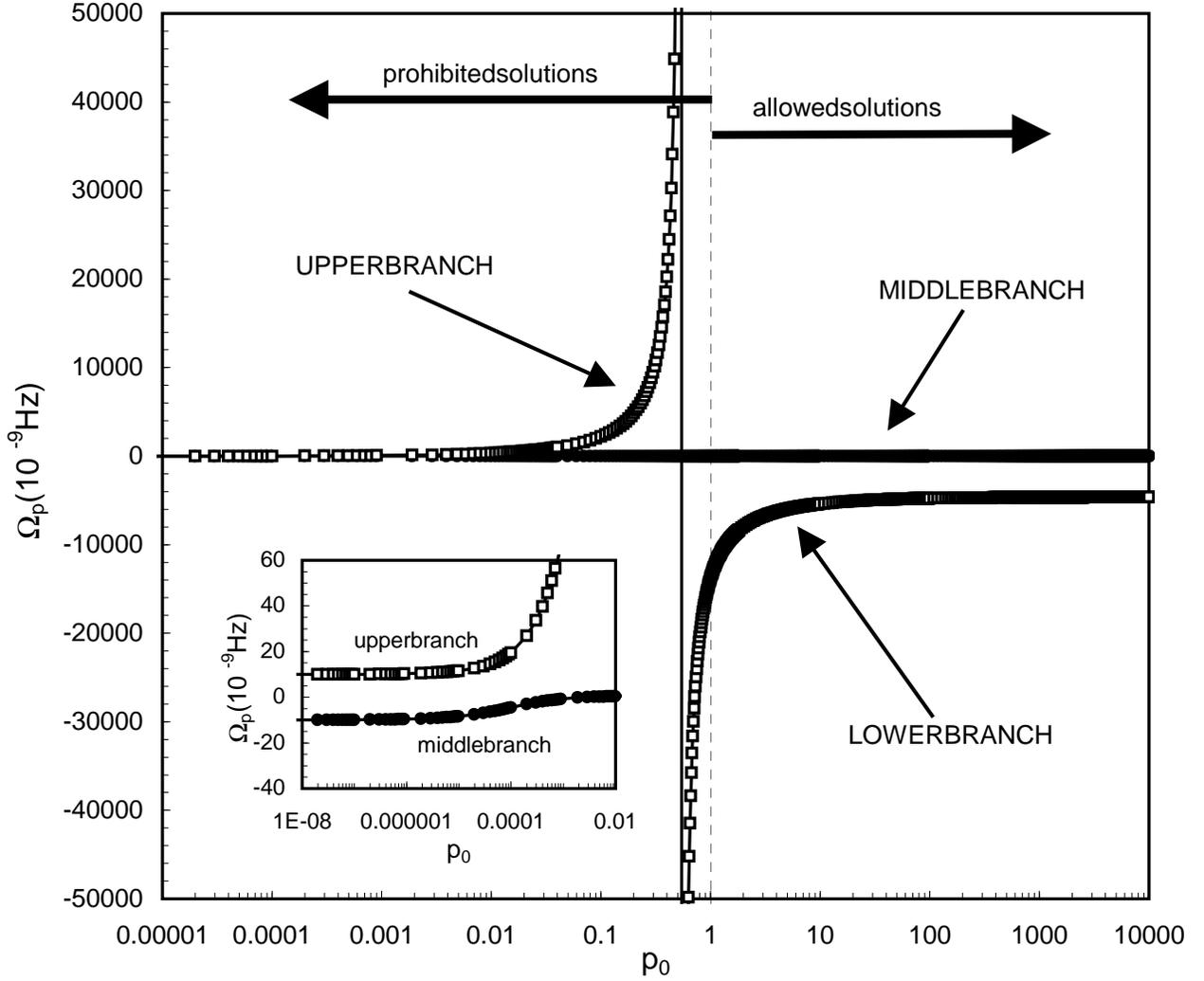}
\caption{Solution for the precession velocity, displaying the three branches. Under the assumption of hydrostatic equilibrium, only the middle and the lower 
branches are allowed ( $p_{0} \geq 1$ ). We have assumed $M_{p} >> M_{s}$, ${\delta} \approx 4 $, and d/r =100}
\label{fig2}
\end{figure}

\clearpage

\begin{figure}
\plotone{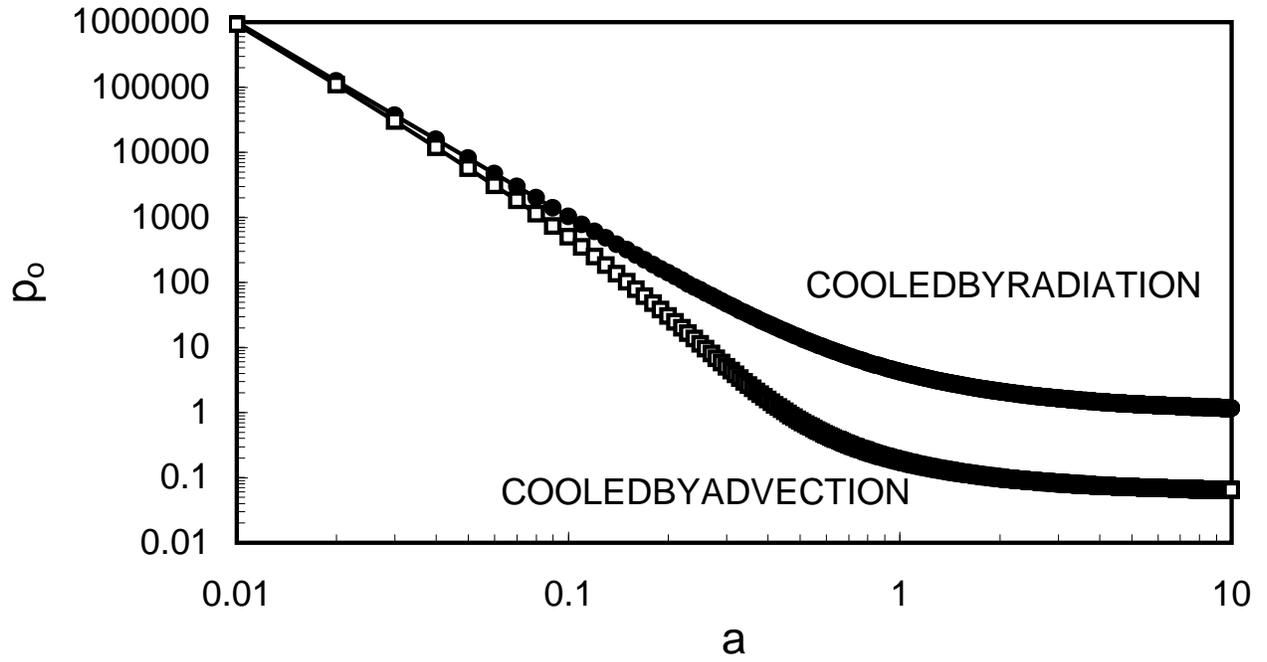}
\caption{$p_0={\left({\ell} \over r \right)}^2$ as a function of $a= {\frac{0.34}{z}} \, \frac{{\dot M}_{1}}{M_{9}} $ for a Schwarzschild Black hole (assumed in the plot), and 
$a= {\frac{1.92}{z}} \, \frac{{\dot M}_{1}}{M_{9}} $ for a maximally rotating Kerr Black Hole. }
\label{fig3}
\end{figure}

\clearpage

\begin{figure}
\plotone{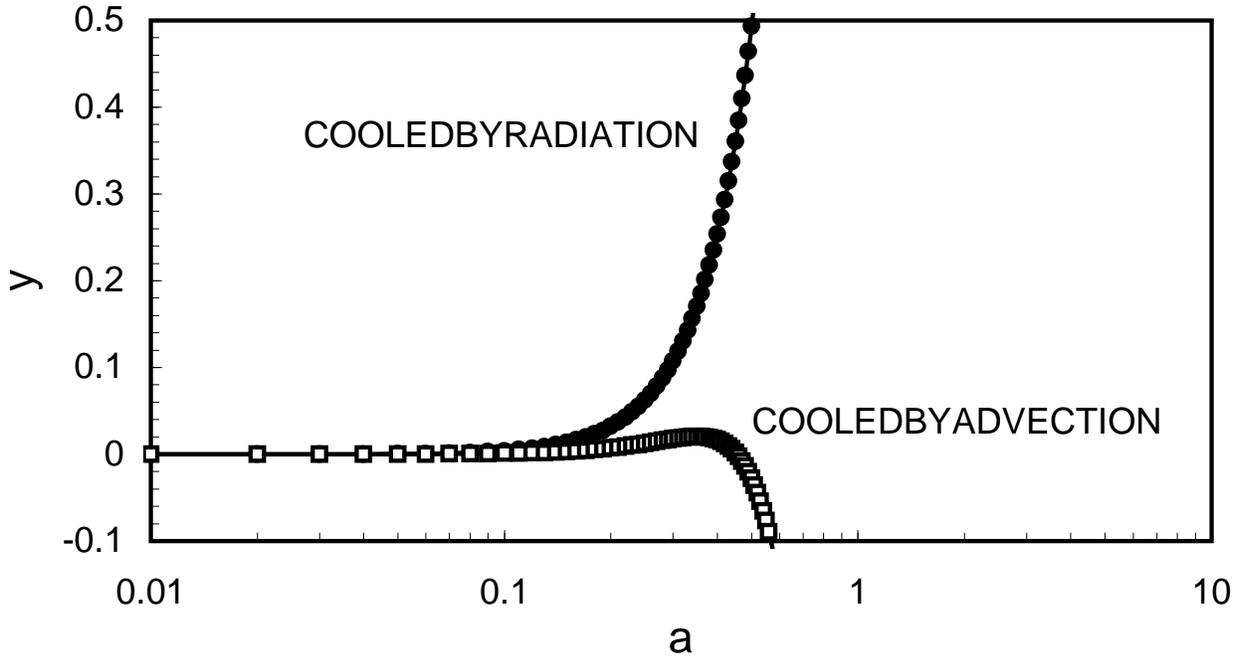}
\caption{y, defined by eq.(100) and eq.(108), as a function of  $a= {\frac{0.34}{z}} \, \frac{{\dot M}_{1}}{M_{9}} $, assuming the primary as a Schwarzshild 
Black Hole} 
\label{fig4}
\end{figure}

\clearpage

\begin{figure}
\plotone{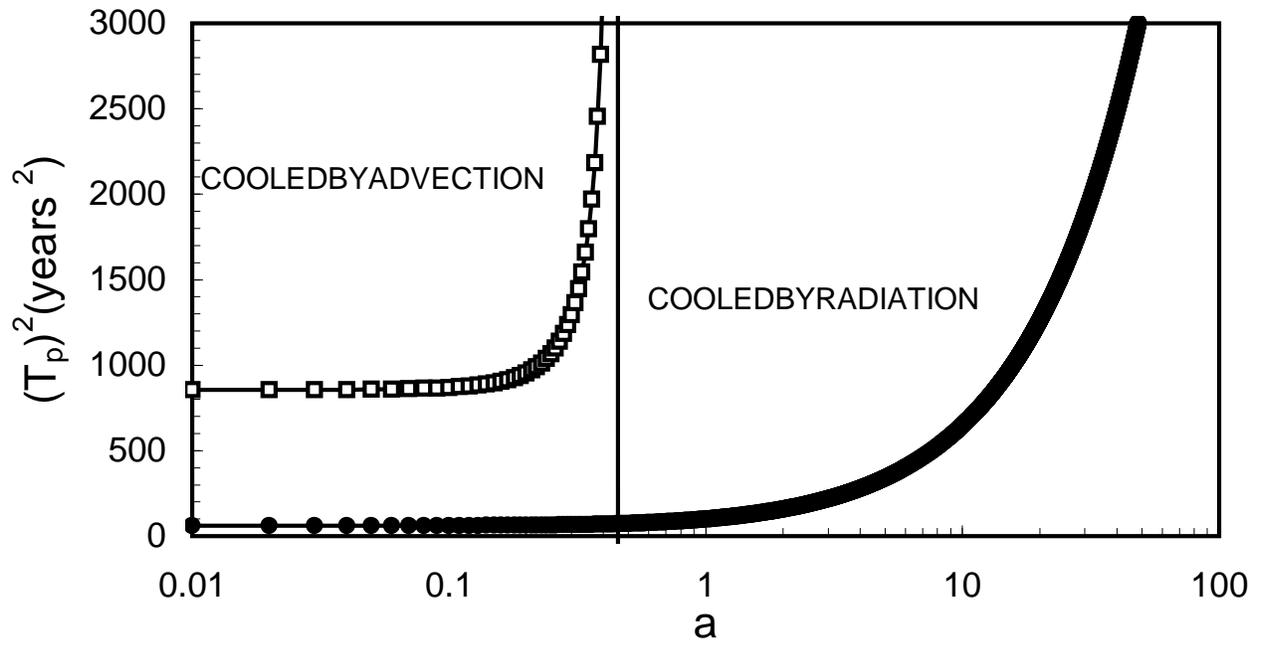}
\caption{The precession period squared ( ${year}^2$) as a function of  $ a= {\frac{0.34}{z}} \, \frac{{\dot M}_{1}}{M_{9}}$, assuming the primary as a Schwarzshild 
Black Hole} 
\label{fig5}
\end{figure}

\end{document}